\documentclass[conference]{IEEEtran}
\ifCLASSINFOpdf
\else
\fi

\usepackage{latexsym}
\usepackage{graphicx}
\usepackage{times}

\begin{document}
\title{\textbf{An Unconditionally Secure Key Management Scheme for Large-Scale Heterogeneous Wireless Sensor Networks}}
\author{\IEEEauthorblockN{Ashok Kumar Das\\ \\}
\IEEEauthorblockA{International Institute of Information Technology, Bhubaneswar\\
Bhubaneswar 751 013, India \\
Email: {\tt iitkgp\_akdas2006@yahoo.co.in}}}

\date{}

\maketitle

\begin{abstract}
Key establishment in sensor networks becomes a challenging problem
because of the resource limitations of the sensors and also due to vulnerability to physical capture of
the sensor nodes.  In this paper, we propose an unconditionally secure probabilistic group-based key pre-distribution scheme for a
 heterogeneous wireless sensor network. The proposed scheme always guarantees that no matter how many sensor nodes are compromised, the non-compromised
nodes can still communicate with $100\%$ secrecy, i.e., the proposed scheme is always unconditionally secure against node capture attacks.
Moreover, it provides significantly better trade-off between communication overhead, computational overhead, network connectivity and security against
node capture as compared to the existing key pre-distribution schemes. It also supports dynamic node addition after the initial deployment of the nodes in the network.
\end{abstract}


\section{Introduction}
In a wireless sensor network a large number of tiny computing nodes, called sensors, are deployed for the purpose of sensing data and then to bring the data back securely to nearby base stations. The base stations then preform the costly computation on behalf of the sensors to analyze the data sensed by the sensors. Due to resource limitations of the nodes and also due to the vulnerability of physical captures of the nodes, the traditional public key cryptographic techniques such as RSA~\cite{jr:01}, Diffie-Hellman key exchange~\cite{jr:02}, El Gamal cryptosystem~\cite{jr:03}, etc. are too much complicated and energy consuming. The symmetric ciphers such as DES, AES, RC5~\cite{bk:04, cf:28} are then the viable options for encrypting/decrypting secret data. In order to use symmetric cipher, we need to establish pairwise keys between communicating sensors. But setting up symmetric keys among communicating nodes remains till now a challenging problem. A survey on sensor networks can be found in \cite{jr:06}.
\par
\indent In order to establish pairwise keys between neighboring sensor nodes, a protocol is used known as the \emph{bootstrapping protocol}. A bootstrapping protocol has the following three phases, called the \emph{key pre-distribution phase}, the \emph{direct key establishment (shared key
 discovery) phase} and the \emph{path key establishment phase}.  Before deployment of nodes in a target field, the key setup server (usually the base station) performs the \emph{key pre-distribution phase}. In this phase each sensor node is loaded by a set of pre-distributed keys in its memory. The next phase occurs immediately after deployment of nodes in the target field. After deployment, the \emph{direct key establishment phase} is performed by nodes in order to establish direct pairwise keys between them. To establish pairwise keys between nodes, each node first discovers its neighbor nodes in its communication range. Two nodes $u$ and $v$ are called \emph{physical neighbors} if they are within communication ranges of one another. In order to discover physical neighbors, each node broadcasts a HELLO message containing its own ID. Thus, each node also receives HELLO message from its neighbor nodes. In this way, each node prepares a list of neighbor nodes which are basically the physical neighbors. Two physical neighbors $u$ and $v$ are called \emph{key neighbors} if they share one or more key(s) in their key rings pre-loaded before deployment during the key pre-distribution phase. Finally, nodes $u$ and $v$ can secretly  and directly communicate with one another if and only if they are both physical and key neighbors. In this case nodes $u$ and $v$ are termed as \emph{direct neighbors}. The final phase known as the \textit{path key establishment phase} is an optional stage and, if executed, adds to the connectivity of the network. Suppose two physical neighbors $u$ and $v$ could not able to establish a pairwise key during the direct key establishment phase because of the fact that they do not share any common key(s) in their key rings. In this phase, a secure path is discovered between $u$ and $v$ and a fresh pairwise key $k$ is sent securely along that path. Thus, nodes $u$ and $v$ use this path key $k$ for their future secret communications. 
\par
\indent Several symmetric key pre-distribution techniques \cite{cf:01, cf:02, cf:04, cf:05, cf:07, jr:10, cf:24, jr:13} are proposed in the literature. Most of these schemes are not scalable and also they are vulnerable to a small number of captured nodes in the network. In this paper, we propose a probabilistic group-based key pre-distribution scheme based on a  heterogeneous wireless sensor network (HWSN). Our scheme makes use of pre-deployment locations of sensors in order to significantly enhance network performances as compared to those for the existing key pre-distribution schemes. 
\par
\indent The rest of the paper is organized as follows. Section II describes briefly the related works. In Section III, we introduce our proposed scheme which is a probabilistic group-based key distribution scheme applied in a heterogeneous wireless sensor network. Section IV gives performance analysis and
security analysis of our scheme. Section V discusses the simulation results of our scheme. In Section VI, we compare the
performances of our scheme with the existing related schemes. Finally, we conclude the paper in Section VII.

\section{Related work}
Eschenauer and Gligor in 2002 first proposed a random key pre-distribution scheme~\cite{cf:01}. Their scheme, henceforth referred to as the EG scheme, consists the following three phases. In the \emph{key pre-distribution phase}, the (key) setup server chooses a pool $\mathcal{K}$ of $M$ randomly generated symmetric keys. Each key is assigned a unique identifier in the pool $\mathcal{K}$. For each sensor node $u$ to be deployed, the setup server picks a random subset $K_{u}$ of size $m$ from the pool $\mathcal{K}$ and loads this subset into its memory. This subset $K_{u}$ is called the \textit{key ring} of the node $u$.  After the sensor nodes are deployed in some target field, a \emph{direct key establishment phase} (also called the shared key discovery phase) is performed by each sensor node in the network. To establish a secret key between them, they exchange the key ids from their key rings in plaintext. If there is a common key id between their key rings, the corresponding key is taken as the secret key between them and they use this key for their future secure communication. Nodes which discover that they have a shared secret key in their key rings then verify that their neighbor actually holds the key through a challenge-response protocol. Since the random subsets for the nodes are drawn from the pool $\mathcal{K}$ randomly without replacement, the same key may be used for secret communication by several pairs of neighbor nodes in the network. The \emph{path key establishment phase} is an optional stage, and if executed, adds to the connectivity of the network. Suppose two neighbor nodes $u$ and $v$ fail to establish a secret key between them in the direct key establishment phase, but there exists a secure path. Once such a secure path is discovered, $u$ generates a new random key $k$ and securely transmits it along this path to the desired destination node $v$. In this way, $u$ and $v$ can communicate secretly and directly using $k$. However, the main problem is that the communication overhead increases significantly with the number $h$ of hops. For this reason, in practice, $h$ is restricted to a small value, say 2 or 3. An improvement of the path key establishment phase has been proposed in \cite{cf:18}, called the key reshuffling scheme, which improves the network performances significantly as compared to those for the path key establishment phase.
\par
\indent The $q$-composite scheme proposed by Chan et al.~\cite{cf:02} is one of the modifications of the EG scheme. In this scheme, two neighbor nodes require at least $q$ common keys $(q > 1)$ instead of one in order to establish a secret key between them. The $q$-composite scheme enhances the security against node capture significantly as compared to that for the EG scheme if the number of captured nodes is small.
\par
\indent In the multipath key reinforcement scheme proposed by Chan et al.~\cite{cf:02}, the main idea is  to strengthen the security of an established link key by establishing the link key through multiple paths. This method can be applied in conjunction with the EG scheme to yield greatly improved resilience against node capture attacks by trading off some network communication overhead.
\par
\indent The random pairwise keys scheme proposed by Chan et al.~\cite{cf:02} is described as follows. Let $m$ be the size of the key ring of each sensor node and $p$ the probability that any two nodes be able to communicate securely. In the key predistribution phase, a total of $n = \frac{m}{p}$ unique node identifiers are generated. The actual size of the network may be smaller than $n$. For each sensor node to be deployed, a set of $m$ other randomly distinct node ids is selected and then a pairwise key is generated for each pair of nodes. The key is stored in both nodes' key rings along with the id of the other node that also knows the key.  In the direct key establishment phase, each node broadcasts its own id to its neighbor nodes in its communication range. Two neighbor nodes can then easily verify the id of a neighbor node in their key rings. If the id of a neighbor node is found in a node's key ring, they share a common pairwise key for communication. A cryptographic handshake is then performed between neighbor nodes for mutual verification of the common key. Since the pairwise key between the two nodes is generated randomly, no matter how many nodes are captured by an adversary, the other non-compromised nodes communicate with each other with $100\%$ secrecy. Thus, the random pairwise keys scheme provides unconditional security against node capture attacks. However, this scheme degrades network connectivity when the network size is large.
\par
\indent The polynomial-based key pre-distribution scheme proposed by Blundo et al. in~\cite{cf:08} is described as follows. In the key pre-distribution phase, an offline key setup server assigns unique identifiers to all the sensor nodes to be deployed in a target field. The setup server then generates randomly a $t$-degree symmetric bivariate polynomial $f(x,y)$, defined by $f(x,y) =$ $\sum_{i,j=0}^{t}\, a_{ij} \, x^{i} \, y^{j}$, where the coefficients $a_{ij}$ $(0 \le i,j \le t)$ are randomly chosen from a finite field $F_q =$ $GF(q)$, $q$ is a prime that is large enough to accommodate a symmetric cryptographic key, with the property that $f(x,y) = f(y,x)$. For each sensor node $u$ to be deployed, the setup server computes a polynomial share $f(u,y)$. We note that $f(u,y)$ is a t-degree univariate polynomial. The setup server finally loads the coefficients of $y^{j}$ of $f(u,y)$ in the memory of the sensor node $u$. In the direct key establishment phase, each sensor node $u$ first locates its physical neighbors in its communication range and broadcasts its own id to its neighbors. Let $u$ and $v$ be two neighbors. After receiving the id of the node $v$, $u$ computes the secret key shared with $v$ as $k_{u,v} = f(u,v)$. Similarly, $v$ computes the secret key shared with $u$ as $k_{v,u} = f(v,u)$. Since $f(u,v) = f(v,u)$, we have $k_{u,v} =$ $k_{v,u}$. Thus, both the nodes $u$ and $v$ store the key $k_{u,v}$ for their future secret communication. The advantage of this scheme is that any two neighbor nodes can establish a secret key using the same symmetric bivariate polynomial $f(x,y)$, and there is no communication overhead during the pairwise key establishment process. The main drawback is that if more than $t$ nodes in the network are compromised by an adversary, he/she can easily reconstruct the original polynomial using \textit{Lagrange interpolation}~\cite{bk:03}. As a result, all the pairwise keys shared between the non-compromised nodes will also be compromised. Thus, this scheme is \textit{unconditionally secure and $t$-collusion resistant}. Although increasing the value of $t$ can improve the security property of this scheme, it is not feasible for wireless sensor networks due to the limited memory in sensors.
\par
\indent Liu and Ning's polynomial-pool based key predistribution scheme~\cite{jr:12} improves security considerably as compared to that for the polynomial-based key pre-distribution scheme, the EG scheme, and the $q$-composite scheme. The location-aware closest pairwise keys scheme (CPKS) based on the random  pairwise keys scheme and closest polynomials pre-distribution scheme (CPPS) based on the polynomial-pool based scheme~\cite{jr:10} improve significantly the performances of network connectivity and resilience against node capture when the deployment error between the actual location and the expected
deployed location of sensor nodes is smaller. The group-based key pre-distribution scheme proposed by Huang et al.~\cite{cf:25} is a matrix based key distribution scheme. Their scheme requires less number of keys preinstalled for each sensor and is resilient to selective node capture attack and node fabrication attack. Liu and Ning proposed a group based key pre-distribution scheme \cite{cf:27} which performs better than the existing schemes \cite{cf:01, cf:02, cf:04}. The deterministic group based key pre-distribution scheme proposed in \cite{cf:26} improves significantly better performances as compared to other existing key pre-distribution schemes \cite{cf:01, cf:02, cf:04, jr:10, cf:24, jr:13}.
\par
\indent The low-energy key management scheme (LEKM)~\cite{cf:24} and improved key distribution mechanism (IKDM) ~\cite{jr:13} are proposed in hierarchical WSNs. These schemes have better performances than the random key distribution schemes~\cite{cf:01,cf:02}, because hierarchical structure has used for those schemes. LEKM requires less key storage overhead than the random schemes~\cite{cf:01,cf:02}. The main drawback of LEKM is that once a cluster head in a cluster is captured, all the keys in sensors of that cluster are compromised. Though IKDM requires only two secret keys to be stored in each sensor's memory, once a cluster head in a cluster is captured after the network initialization phase, all the keys stored in sensors in that cluster are compromised. The basic problem in LEKM and IKDM is that all the sensors in a cluster communicate directly with the cluster head only.

\section{The proposed scheme}
In this section, we first describe in brief the network model used for developing our scheme. We then describe the main motivation behind development of our scheme. Finally, we describe our proposed scheme.

\subsection{Network Model}
In this section, we discuss a heterogeneous network model which will be used for development of our proposed scheme.

\begin{figure}[htbp]
\centering
\includegraphics[scale=0.49]{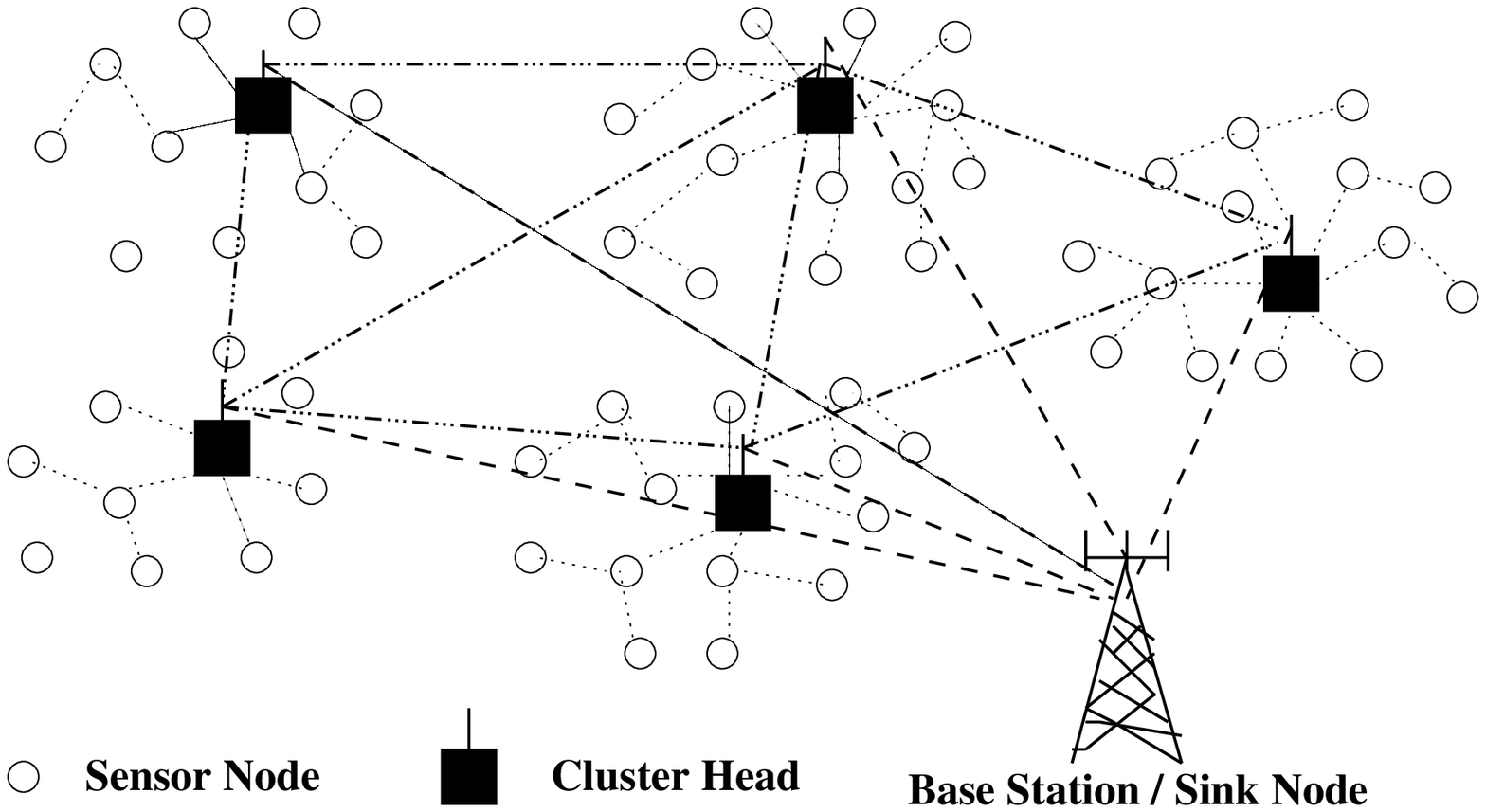}
\caption{A heterogeneous wireless sensor network (HWSN) architecture.} \label{Figure:}
\end{figure}

\par
\indent   A \textit{heterogeneous wireless sensor network (HWSN)} is shown in Figure 1. From this figure, we see that there is a hierarchy among the nodes based on their capabilities: \emph{base station}, \emph{cluster heads} and \emph{sensor nodes}. \textit{Sensor nodes} are inexpensive, limited capability and generic wireless devices. Each sensor has limited battery power, memory size and data processing capability and short radio transmission range. Sensor nodes in a group (also called a cluster) communicate among each other in that cluster and finally communicate with the cluster head (CH). \textit{Cluster heads} have more resources than sensors. They are equipped with high power batteries, larger memory storage, powerful antenna and data processing capabilities. Cluster heads can execute relatively complicated numerical operations than sensors and have much larger radio transmission range. Cluster heads can communicate with each other directly and relay data between its cluster members and the base station. A \textit{base station} or \textit{sink node} (BS) is typically a gateway to another network, a powerful data processing/storage center, or an access point for human interface.  A base station collects sensor readings, performs costly operations on behalf of sensor nodes and manages the network. In some applications, the base station is assumed to be trusted. Thus, the base station is used as key distribution center (KDC). 
\par
\indent Sensor nodes are deployed around one or more hop neighborhood of the base station. Since the base station is most powerful node in the network, it can reach all the sensor nodes  in that network. Depending on the applications, the base station (BS) can be located either in the center or at a
corner of the network. Data flow in such networks can be: $(i)$ pairwise (unicast) among sensor nodes, $(ii)$  group-wise (multicast) within a cluster of sensor nodes, and $(iii)$ network-wise (broadcast) from base station to sensor nodes.

\subsection{Motivation}
Our scheme is motivated by the followings. In many sensing applications, connectivity between all sensor nodes is not necessary. Thus, data centric mechanism should be performed to aggregate redundant data in order to reduce the energy consumption and traffic load in wireless sensor networks. Therefore, the heterogeneous network model has more operational advantages over the distributed homogeneous model for wireless sensor networks due to inherent limitations of sensors on power and processing capabilities.
\par
\indent The random pairwise keys scheme \cite{cf:02} has the following limitations. Though this scheme always provides unconditional security against node capture, it provides very low network connectivity in particularly when the network size is large. In practice, the sensor network is assumed to be highly scalable and hence the random pairwise keys scheme is not applicable in large-scale distributed sensor networks.
\par
\indent The group-based deterministic key distribution mechanism \cite{cf:26} based on bivariate polynomials provides very high network connectivity and unconditional security against node capture. But this scheme requires computational overhead due to evaluation of a $t$-degree polynomial over a finite field $F_q$. In this paper, we propose an energy efficient key distribution scheme. Our scheme is an improved version of this group-based deterministic key distribution mechanism \cite{cf:26} based a heterogeneous network model (as shown in Figure 1) which requires significantly low computational and communication overheads in order to establish pairwise secret keys between communicating nodes in a sensor network. 

\subsection{Our approach}
As in \cite{cf:26}, we consider a heterogeneous wireless sensor network (HWSN) consisting of two types of sensors: a small number of powerful High-end sensors (H-sensors) and a large number of resource-constrained Low-end sensors (L-sensors). H-sensors can execute relatively complicated numerical operations than L-sensors and have much larger radio transmission range and larger storage space than L-sensor nodes. On the other hand, L-sensors are extremely resource-constrained. For example, the H-sensors can be PDAs and the L-sensors are the MICA2-DOT motes~\cite{ms:03}. We also assume that the target field is two dimensional and partitioned into a number $l$ of equal sized disjoint groups (clusters). Each group will consist of a group head $GH_{i}$ (here it is an H-sensor node) and a number $n_i$ of L-sensor nodes. The number $n_i$ of regular sensor nodes is to be taken in each deployment group so that the network connectivity in each group is reasonably high. L-sensors are to be deployed randomly in a group only and each group head will be deployed in that group around the center of that group. For our sake of simplicity, we call an L-sensor node as regular sensor node. The base station (BS) can be located either in the center or at a corner of the network.
\par
\indent The following assumptions are made while constructing our protocol.
\begin{itemize}
\item After deployment of the nodes in a target field, each L-sensor (regular sensor node) as well as H-sensor nodes (group heads) are assumed to be static only.
\item Base station is assumed to be trusted and it will never be compromised by an attacker.
\item An adversary can eavesdrop on all traffic, inject packets and reply old messages previously delivered. If an adversary captures a node, all the keying information it holds will also be compromised.
\end{itemize}

\indent Our scheme makes use of the existing polynomial-based key pre-distribution scheme in order to establish pairwise keys among group heads in a sensor network. We use the extended version of the random pairwise keys scheme in order to facilitate establishment of pairwise keys among regular sensor nodes in a group.  
\par

\indent Our scheme consists of the following phases.

\subsubsection{Key pre-distribution phase}
This phase is performed by the (key) setup server in offline before deployment of the sensor nodes in a target field. The steps involved in this phase are as follows:

\begin{itemize}
\item \textit{Step-1: }The setup server first assigns a unique identifier, say  $id_{GH_i}$ to each group head $GH_i$ which will be deployed in the target field. For each deployed regular sensor node $u$, the setup server also assigns a unique identifier, say $id_u$.

\item \textit{Step-2: }The setup server then selects randomly a unique master key, say $MK_{GH_i}$ for each group head $GH_i$. This master key is shared between the group head $GH_i$ and the base station only. The setup server also assigns for each deployed regular sensor node $u$ a unique randomly generated master key, say $MK_u$ which is shared with the base station only.

\item \textit{Step-3: }For each deployment group $G_i$, the setup server generates a node pool, say $N_i$ consisting of the IDs of the group head $GH_i$ and the $n_i$ regular sensor nodes to be deployed in that group.

\item \textit{Step-4: }For each deployed regular sensor node $u$ in each group $G_i$, the setup server selects a set $S_i$ consisting of randomly chosen $m$ node IDs from the corresponding node pool $N_i$ of that group $G_i$. Let the set $S_i$ be as $S_i =$ $\{id_{v_1}, id_{v_2}, \ldots, id_{v_m} \}$. We note that one of the IDs in $S_i$ may be the ID of the group head $GH_i$. Then for each pair $(u, v_j)$, $(j = 1, 2, \ldots, m)$, the setup server computes the $m$ key-plus-id combinations, say $\{ (SK_{u,v_j}, id_{v_j}), j = 1, 2, \ldots, m \}$, where $SK_{u, v_j}$ $ = PRF_{MK_{v_j}}(id_u)$. Here $PRF$ is a pseudo random function proposed by Goldreich et al.~\cite{jr:08}.

\item \textit{Step-5: }For all the $m$ deployed group heads $GH_i$ $(i = 1,$ $2,$ $\ldots,$ $m)$, the setup server randomly generates a $t$-degree bivariate polynomial $f(x,y)$ $\in F_q[x,y]$ over a finite field $F_q$, with the property that $f(x,y) =$ $f(y,x)$, that is, $f(x,y)$ is symmetric such that $t >> l$. The reason for choosing the degree of the polynomial $f(x,y)$ to be higher is that even if an adversary captures all the $l$ group heads in the network, the polynomial $f(x,y)$ will never be compromised. The setup server then computes a polynomial share $f(id_{GH_i},y)$ for each deployed group head $GH_i$ $(i = 1,$ $2,$ $\ldots,$ $m)$.

\item \textit{Step-6: }Since the group heads are H-sensors and are more powerful nodes than regular sensor nodes, we can store more keying information in their memory. For each deployed group head $GH_i$ $(i = 1,$ $2,$ $\ldots,$ $l)$, the setup server randomly selects a set $S$ $= \{id_{w_1}, id_{w_2},\ldots, id_{w_{m'}} \}$ from the node pool $N_i$ corresponding to that group $G_i$, where $m' \ge m$. Then for each pair $(GH_i, w_j)$, $(j = 1, 2, \ldots, m')$, the setup server also computes the $m'$ key-plus-id combinations, say $\{ (SK_{GH_i,w_j}, id_{w_j}), j = 1, 2, \ldots, m' \}$, where $SK_{GH_i, w_j}$ $ = PRF_{MK_{w_j}}(id_{GH_i})$.

\item \textit{Step-7: }Finally, the setup server loads the following information into the memory of each group head $GH_i$ $(i = 1,$ $2,$ $\ldots,$ $l)$: $(i)$ its own identifier, $(ii)$ its own master key $MK_{GH_i}$, $(iii)$ the polynomial share $f(id_{GH_i}, y)$ computed in step-5, and $(iv)$ $m'$ key-plus-id combinations computed in step-6. Each deployed regular sensor node $u$ in the deployment group $G_i$ is loaded with the following information: $(i)$ its own identifier, $(ii)$ its own master key $MK_u$,  and $(iii)$ $m$ key-plus-id combinations computed in step-4. The loaded information in each regular sensor node as well as group head are shown in Tables I and II.
\end{itemize}

\begin{table}[htbp]
\caption{Key ring of a regular sensor node $u$ in its deployment group $G_i$} \label{Table: } 
\begin{center}
\begin{tabular}{|c|} \hline 
$id_u$ \\ \hline
$MK_{u}$ \\ \hline
$\{ (SK_{u,v_j}, id_{v_j}), j = 1, 2, \ldots, m \}$, \\
 $SK_{u, v_j}$ $ = PRF_{MK_{v_j}}(id_u)$  \\ \hline
\end{tabular}
\end{center}
\end{table}

\begin{table}[htbp]
\caption{Key ring of a group head $GH_i$ in its deployment group $G_i$} \label{Table: } 
\begin{center}
\begin{tabular}{|c|} \hline 
$id_{GH_i}$ \\ \hline
$MK_{GH_i}$ \\ \hline
$f(id_{GH_i},y)$ \\ \hline
$\{ (SK_{GH_i,w_j}, id_{w_j}), j = 1, 2, \ldots, m' \}$, \\
 $SK_{GH_i, w_j}$ $ = PRF_{MK_{w_j}}(id_{GH_i})$  \\ \hline
\end{tabular}
\end{center}
\end{table}

\indent We note that a typical regular sensor node can store $200$ keys in its memory. Hence we take the value of $m$ as $m = 200$, whereas the value of $m'$ will be taken larger than $m$ due to large storage memory of group heads.

\subsubsection{Direct key establishment phase}
As soon as regular sensor nodes are deployed randomly in their respective groups, their first task is to locate the physical neighbors within their communication ranges. Group heads in their groups locate their physical neighbors which are the regular sensor nodes. Group heads also locate their other group heads in their communication ranges in the network. 
\par
\indent In our direct key establishment phase, we have the following two pairwise key establishment procedures: one is the inter-group pairwise key establishment and other is the intra-group pairwise key establishment. In the inter-group pairwise key establishment, only group heads will establish pairwise secret keys with their neighbor group heads. On the other hand, during the intra-group pairwise key establishment the regular sensor nodes will establish pairwise keys with their neighbor nodes in their own deployment group, and also the group heads will establish pairwise keys with their neighbor regular sensor nodes in their own deployment group.\\

\noindent \textit{(a) Inter-group pairwise key establishment} \\
\indent If $GH_i$ and $GH_j$ be two neighbor group heads, they can establish pairwise secret key by exchanging their own ids $id_{GH_i}$ and $id_{GH_j}$. After exchanging their ids, $GH_i$ computes the pairwise secret key as $f(id_{GH_i},id_{GH_j})$ by just evaluating its own polynomial share $f(id_{GH_i},y)$ at the point $y =$ $id_{GH_j}$. In a similar fashion, $GH_j$ computes a secret key $f(id_{GH_j},id_{GH_i})$ by evaluating its polynomial share $f(id_{GH_j},y)$ at the point $y =$ $id_{GH_i}$. Since the polynomial is symmetric, so the shared secret key between the group heads $GH_i$ and $GH_j$ is  $SK_{GH_i,GH_j} = $ $f(id_{GH_i},id_{GH_j})$. Finally, they store this key $SK_{GH_i,GH_j}$ for their future secure communication.\\

\noindent \textit{(b) Intra-group pairwise key establishment} \\
\indent In this phase, we consider the following three cases:\\

\noindent \textit{Case I: regular node to regular node key establishment}\\
\indent In order to establish a secret pairwise key between two neighbor regular sensor nodes, say $u$ and $v$ in a deployment group $G_i$, they exchange their own ids $id_u$ and $id_v$. Let the ID of node $v$ be resident in the key ring of node $u$. Then from Table I, we note that $u$ is sharing a pairwise key with node $v$. Node $u$ then informs node $v$ that it is sharing a pairwise key $SK_{u,v}$. This notification contains the ID of node $u$ with a small request message. It is noted that this notification never contains the exact key $SK_{u,v}$. After receiving the request from $u$, node $v$ can easily compute the same pairwise key $SK_{u,v}$ by computing PRF function with the help of its own master key $MK_v$ and the ID of node $u$ as $SK_{u,v}$ $= PRF_{MK_{v}}(id_u)$. Node $v$ then stores this key $SK_{u,v}$ for future secret communication with the node $u$.\\

\noindent \textit{Case II: group head to regular node key establishment}\\
\indent In order to establish a secret key between a regular sensor node $u$ and its group head $GH_i$ which is within its communication range, they need to exchange their own ids. If the ID of node $u$ is resident in the key ring of the group head $GH_i$, then it informs to $u$ that it has a pairwise key shared with $u$. This is done by sending a short notification containing the ID of $GH_i$ to node $u$. After receiving this notification, $u$ can easily compute the shared secret pairwise key with $GH_i$ as $SK_{GH_i, u} =$ $PRF_{MK_{u}}(id_{GH_i})$ and store this key for future communication with $GH_i$. Now, if the ID of $u$ is not resident in the key ring of $GH_i$, it is also possible that the ID of $GH_i$ is resident in the key ring of node $u$. In this case, $u$ sends a short notification containing its own ID to group head $GH_i$. Then $GH_i$ computes the shared secret pairwise key $SK_{GH_i,u}$ with $u$ as $SK_{GH_i, u} = $ $PRF_{MK_{GH_i}}(id_{u})$ using its own master key and the ID of node $u$. $GH_i$ then stores this key for future secret communication with node $u$.\\

\noindent \textit{Case III: regular node to regular node key establishment with help of another group head}\\
\indent This is a spacial case considered here. Assume that a regular node was supposed to be deployed in its group $G_i$. But due to some deployment error during deployment, it is deployed to some other group, say $G_j$. It is then noted that $u$ could not able to establish secret keys with its neighbor regular nodes in that group because it does not have any keying information containing in that group. Therefore, we need for the node $u$ to establish pairwise keys with its neighbor nodes with the help of the group head $GH_j$ in $G_j$ as follows (as in \cite{cf:26}).
\par
\indent In order to establish a pairwise key between $u$ and its neighbor node $v$, node $u$ sends a request containing of its own id $id_u$ and a randomly generated nonce $RN_u$. After receiving such a request, node $v$ generates a random nonce $RN_v$ and sends a request consisting of its own id $id_v$ as well as the id of $u$, $id_u$, random nonces $RN_u$ and $RN_v$ to its own group head $GH_j$ which is protected by its own master key $MK_v$. Then the group head $GH_j$ forwards this request to its neighbor group head and finally this request comes eventually to the base station. The base station first validates this request by decrypting the request by the master key $MK_v$ of the node $v$, because the base station has the master key $MK_v$ of $v$. If the validation passes, the base station then only generates a secret random key $k_{u,v}$ to be shared by the nodes $u$ and $v$. Then it makes two protected copies: one for node $u$, $E_{MK_u}(k_{u,v}\oplus id_u \oplus RN_u)$ and other for node $v$, $E_{MK_v}(k_{u,v}\oplus id_v \oplus RN_v)$ where $E_{k}(M)$ denotes the encryption of data $M$ using the key $k$. The first one is sent to node $u$ and the later copy is sent to node $v$ via group heads. Nodes $u$ and $v$ first decrypt their protected copies. Node $u$ retrieves the secret key $k_{u,v}$ using its own id and its own random nonce $RN_u$ as $k_{u,v} =$ $(k_{u,v}\oplus id_u \oplus RN_u)$ $\oplus (id_u \oplus RN_u)$. Similarly, node $v$ also uses its own id  and random nonce $RN_v$ in order to retrieve the secret key $k_{u,v}$ as $k_{u,v} =$ $(k_{u,v}\oplus id_v \oplus RN_v)$ $\oplus (id_v \oplus RN_v)$.  We also note that the communication overhead is not much due to involvement of the group heads during this process. In fact, such a scenario is unlikely to occur, because the probability of having a smaller deployment error is typically higher than the probability of having a larger one when the nodes are randomly deployed in a deployment group. In a similar fashion, node $u$ can also establish a secret key with the group head $GH_j$ if $GH_j$ is neighbor of $u$.

\subsubsection{Dynamic sensor node addition phase}
In order to add a new regular sensor node $u$ in a particular deployment group, say $GH_i$, the key setup server assigns a unique id, say $id_u$ and randomly generates a master key $MK_u$ for $u$ which will be shared with the base station only. Then the setup server selects a set $S_i$ consisting of randomly chosen $m$ node IDs from the corresponding node pool $N_i$ of that group $G_i$. Let the set $S_i$ be as $S_i =$ $\{id_{v_1}, id_{v_2}, \ldots, id_{v_m} \}$. We note that one of the IDs in $S_i$ may be the ID of the group head $GH_i$. Then for each pair $(u, v_j)$, $(j = 1, 2, \ldots, m)$, the setup server computes the $m$ key-plus-id combinations, say $\{ (SK_{u,v_j}, id_{v_j}), j = 1, 2, \ldots, m \}$, where $SK_{u, v_j}$ $ = PRF_{MK_{v_j}}(id_u)$ and loads these information in its memory.  
\par
\indent After deployment in its own deployment group, it establishes secret keys with its neighbor nodes within its group as described in the \textit{intra-group pairwise key establishment phase}. 

\subsubsection{Dynamic group-head addition phase}
We now consider that a group head $GH_i$ in a group $G_i$ is captured by an adversary. Thus, we need to add a new
group head, say, $GH_i^{'}$ in that group $G_i$ in order to replace that node $GH_i$. In order to add the group head $GH_i^{'}$, the setup server assigns a unique id, say  $id_{GH_i^{'}}$ and a randomly generated  master key $MK_{GH_i^{'}}$ which will be shared with the base station only. The setup server then randomly selects a set $S$ $= \{id_{w_1}, id_{w_2},\ldots, id_{w_{m'}} \}$ from the node pool $N_i$ corresponding to that group $G_i$, where $m' \ge m$. Then for each pair $(GH_i^{'}, w_j)$, $(j = 1, 2, \ldots, m')$, the setup server also computes the $m'$ key-plus-id combinations, say $\{ (SK_{GH_i^{'},w_j}, id_{w_j}), j = 1, 2, \ldots, m' \}$, where $SK_{GH_i^{'}, w_j}$ $ = PRF_{MK_{w_j}}(id_{GH_i^{'}})$. The setup server loads the following information in its memory: $(i)$ the identifier $id_{GH_i^{'}}$ for $GH_i^{'}$, $(ii)$ randomly generated master key $MK_{GH_i^{'}}$, $(iii)$ the polynomial share $f(id_{GH_i^{'}},y)$, and $(iv)$ $m'$ key-plus-id combinations as computed above.
\par
\indent After deployment in the group $G_i$, the group head $GH_i^{'}$ establishes pairwise keys with its neighbor group heads using the \textit{inter-group pairwise key establishment phase} and with the regular sensor nodes using the \textit{intra-group pairwise key establishment phase}.

\section{Analysis of our scheme}
In this section, we analyze the network connectivity of our scheme which is the probability that any two neighbor nodes in a deployment group can establish a secret pairwise key between them. We then discuss the resilience against node capture of our scheme. Finally, we analyze the overhead requirements for storage, communication and computation for key establishment between two neighbor regular sensor nodes.

\subsection{Network connectivity}
From inter-group pairwise key establishment phase described in Section III.C.2, we note that every group head can establish a pairwise secret key with its neighbor group heads in the network using its own polynomial share. Let $p_{grouphead-grouphead}$ denote the probability that a group head can establish a pairwise secret key with its another neighbor group head. Then, we have,
\begin{equation}
p_{grouphead-grouphead} = 1.
\end{equation} 

\indent Now, we will concentrate on the network connectivity in each deployment group $G_i$ $(i=1, 2, \ldots, l)$. Let us first consider the case where a regular sensor node $u$ can establish a pairwise key with its another neighbor regular sensor node $v$ in their group $G_i$. From intra-group pairwise key establishment phase described in Section III.C.2, we see that $u$ and $v$ can establish a pairwise key if any one of the following two events occur: \\\noindent $E_1:$ the event that the ID of node $u$ is resident in $v$'s key ring \\
\noindent $E_2:$ the event that the ID of node $v$ is resident in $u$'s key ring
\par
\indent Let $p_1$ denote the probability that the id of a node will be resident in another node's key ring. Then we have $p_1 = $ $P(E_1) = $ $P(E_2)$. The total number of ways to select $m$ ids from the pool $N_i$ of size $n_i + 1$ is $ {n_i + 1\choose m }$. For a fixed key ring of node $u$, the total number of ways to select key ring of a node $v$ such that key ring of $v$ does not have the id of $u$ is $ {(n_i+1)-1\choose m} $ $= {n_i\choose m}$.  Thus, we have, 
\begin{equation}
p_{1} = \left\{ \begin{array}{c}
                1 - \frac{{n_i\choose m}}
                      {{n_i+1\choose m}} 
           = \frac{m}{n_i+1}, ~$if$ ~m < n_i + 1. \\
             1, ~$if$ ~m \ge n_i + 1.
                \end{array}
        \right.
\end{equation}

\noindent Let $p_{sensor-sensor}$ be the probability that two neighboring regular sensor nodes $u$ and $v$ can establish a pairwise key in a group $G_i$. Then we have, $p_{sensor-sensor}$ $ =  1 - $ (probability that none of $u$ and $v$ will establish a pairwise key). Hence, 
\begin{eqnarray}
p_{sensor-sensor} & = & 1 - (1 - p_1)^{2}.
\end{eqnarray}

\indent We now consider the probability of establishing a pairwise key between a group head $GH_i$ and its neighboring regular sensor node $u$ in a group $G_i$. Let $p_2$ be the probability that the id of $u$ will be resident in key ring of $GH_i$. Then it is easy to deduce (as derived for $p_1$) that
\begin{equation}
p_{2} = \left\{ \begin{array}{c}
                1 - \frac{{n_i\choose m'}}
                      {{n_i+1\choose m'}} 
           = \frac{m'}{n_i+1}, ~$if$ ~m' < n_i + 1. \\
             1, ~$if$ ~m' \ge n_i + 1.
                \end{array}
        \right.
\end{equation}

\noindent If $p_{grouphead-sensor}$ represents the probability that a key is established between $GH_i$ and $u$ in group $G_i$, we have
\begin{equation}
p_{grouphead-sensor} =  1 - (1 - p_1) (1 - p_2).
\end{equation}
\par
\noindent \textit{Overall network connectivity in a group $G_i$: }We note that each group $G_i$ contains at most $n_i$ regular sensor nodes and a group head $GH_i$. Thus, $\mid G_i \mid = n_i + 1$. Let each node have $d$ average number of neighbor nodes. We consider each group is an undirected graph having $n_i + 1$ nodes, each node having the degree $d$. Then the total direct communication links in the group becomes the total number of edges in $G_i$ which is equal to $\frac{(n_i + 1) d}{2}$. The total number of secure direct links formed in the group $G_i$ by the regular sensor nodes and the group head $G_i$ are $\frac{n_i\times d}{2} \cdot p_{sensor-sensor}$ and $d\cdot p_{grouphead-sensor}$ respectively. Thus, we have  $\frac{n_i\times d}{2} \cdot p_{sensor-sensor}$ $+ d\cdot p_{grouphead-sensor}$ secure links out of the total $\frac{(n_i + 1)\times d}{2}$ direct links. Hence, the overall network connectivity in $G_i$ can be estimated as
\begin{eqnarray}
p_{overall}  =  \frac{\frac{n_i\times d}{2} \cdot p_{sensor-sensor} + d\cdot p_{grouphead-sensor}}{((n_i + 1)\times d)/2} \nonumber \\
=  \frac{n_i\cdot p_{sensor-sensor} + 2 \cdot p_{grouphead-sensor}}{n_i + 1}.
\end{eqnarray}

\begin{figure}[htbp]
\centering
\includegraphics[scale=0.33,angle=270]{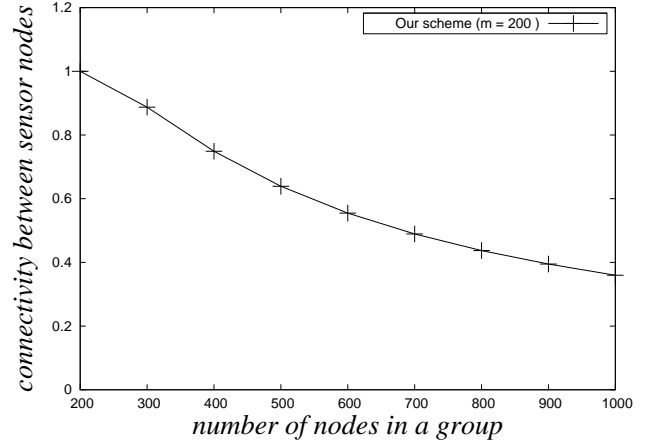}
\caption{Number of regular sensor nodes vs. network connectivity between regular sensor nodes in a group $G_i$, with $m = 200$.} \label{Figure: }
\end{figure}

\begin{figure}[htbp]
\centering
\includegraphics[scale=0.33,angle=270]{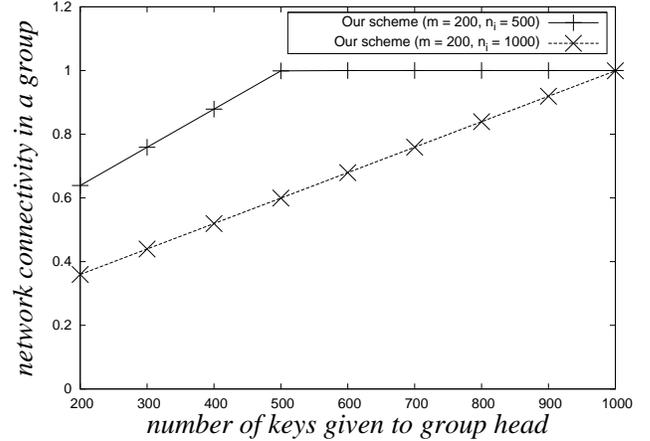}
\caption{Network connectivity between a regular sensor node and its group head $GH_i$ in a group $G_i$, with $m = 200$, $n_i = 500$, $1000$, and different values of $m'$.} \label{Figure: }
\end{figure}

\indent Figure 2 shows the relationship between the network connectivity among regular sensor nodes and the number of nodes in a group. We assume that each regular sensor node is capable of holding $200$ cryptographic keys in its memory (i.e., $m = 200$). It is clear to see from this figure that network connectivity increases when the number of regular sensor nodes in group is smaller. We also note that even if the number of regular sensor nodes reaches $1000$, the network connectivity between regular sensor nodes in that group remains high.
\par
\indent Figure 3 illustrates the network connectivity among a group head and its neighbor regular sensor node in a group. Since the group head is powerful node than regular sensors, loading of an excessive amount of keying materials gives very high network connectivity between that group head and its neighbor regular sensor node.

\subsection{Resilience against node capture}
The resilience against node capture attack of a key distribution scheme is measured
by estimating the fraction of total secure communications that are
compromised by a capture of $c$ nodes \textit{not including} the
communication in which the compromised nodes are directly
involved. In other words, we want to find out the probability that the adversary can
decrypt the secret communications between two non-compromised nodes $u$ and $v$ when $c$
 sensor nodes are already compromised.
\par
\indent From our direct key establishment phase, we notice that each group head $GH_i$ is given a $t$-degree polynomial share $f(id_{GH_i},y)$ for  establishing pairwise keys with its neighbor group heads and the degree of this polynomial is greater than the total number of group heads in the network. The pairwise keys established by the group heads are different. Based on the security of the polynomial-based key pre-distribution scheme \cite{cf:08} even if an adversary captures all the group heads, he/she could not able to compromise this polynomial.
\par
\indent Based on the security of the PRF function~\cite{jr:08}, if a node's master key is not disclosed, no matter how many pairwise keys generated by this master key are disclosed, the task is still computationally difficult for an adversary to recover the master key  as well as the non-disclosed pairwise keys generated with different ids of sensor nodes. Since each pre-distributed pairwise key between two regular sensor nodes, and a regular sensor node and its group head are generated using PRF function randomly, no matter how many nodes are captured, the direct pairwise keys between non-captured nodes are still secure. In other words, node compromise does not eventually lead to compromise of direct pairwise keys between other non-captured nodes, that is, any two non-captured neighboring nodes communicate with $100\%$ secrecy. Hence, our scheme is always unconditionally secure against node capture attack.

\subsection{Overheads}
In this section, we only consider overheads required by the regular sensor nodes, because they are resource-constrained.
\par
\indent From the key pre-distribution phase (described in Section III.C.1) we see that every regular sensor node requires to store its own master key as well as $m$ key-plus-id combinations in its memory. Thus, the storage overhead is mainly due to storing $m+1$ keys. 
\par
\indent A regular sensor node in a deployment group needs to exchange a short request message containing its own id with its neighbor node in that group in order to establish a pairwise key between them, if the id of the neighbor node is resident in its key ring. For the special case described in the direct key establishment phase in Section III.C.2, if a regular node which was expected to deploy in a group but during deployment it is deployed in another group, it requires to establish a pairwise key with its neighbor nodes in that group with the help of group heads. Since the probability of having a smaller deployment error is typically higher than the probability of having a larger one when the nodes are randomly deployed in a deployment group, such a situation is unlikely to occur frequently. Thus, the communication overhead is mainly due to transmission of a short request message.
\par
\indent In order to establish a pairwise key, a regular sensor node needs to perform a PRF operation. Zhu et al. \cite{jr:17} pointed out due to the computational efficiency of pseudo random functions, the computational overhead of the PRF function is negligible. Hence, the computational overhead of our scheme is low as compared to that of computation of a $t$-degree polynomial over a finite field $F_q$ as in \cite{cf:08, cf:09, cf:26}.

\section{Simulation Results}
In this section, we discuss the simulation results of network connectivity in each group. 
\par
\indent We have implemented our scheme in C. We have taken a square deployment field for our simulation.  The target field is partitioned into $l$ groups $G_i$ $(i = 1, 2, \ldots, l)$, each of equal size. For each group $G_i$, we have deployed a group head $GH_i$ around the center of the group. The number $n_i$ of regular sensor nodes is taken to be equal for each group. We deploy the $n_i$ regular sensor nodes randomly in each group $G_i$. The following parameters are considered for our simulation:
\begin{itemize}
\item The number of groups in the target field is $l = 100$.
\item The number of regular sensor nodes deployed in each group is $\le 1000$.
\item The area of the deployment field is $A = 1000m \times 1000m$.
\item The area of each group is $100m \times 100m$.
\item The communication range of each regular sensor node is $30$ meters.
\item The average number of nodes for each node is $\le 100$.
\end{itemize}  

\noindent We have simulated overall network connectivity for each group and then taken the average overall network connectivity for a group. Figures 4 and 5 show the relationship between the simulated overall network connectivity in a group versus the analytical overall network connectivity in that group, with $m = 200$, and different values of $m'$. We observe that both the simulation as well as analysis results tally closely.  
\begin{figure}[htbp]
\centering
\includegraphics[scale=0.33,angle=270]{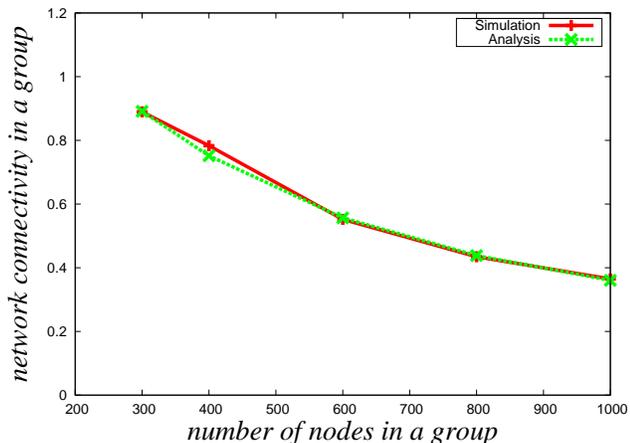}
\caption{Average overall network connectivity of a group $G_i$, with $m = 200$ and $m' = 200$.} \label{Figure: }
\end{figure}

\begin{figure}[htbp]
\centering
\includegraphics[scale=0.33,angle=270]{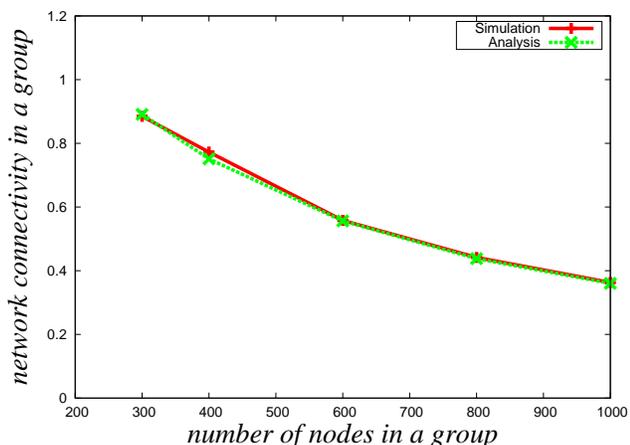}
\caption{Average overall network connectivity of a group $G_i$, with $m = 200$ and $m' = 300$.} \label{Figure: }
\end{figure}

\section{Comparison with previous schemes}
In this section, we compare security against node capture of our scheme with that for the existing schemes.
\begin{figure}[h]
\centering
\includegraphics[scale=0.33,angle=270]{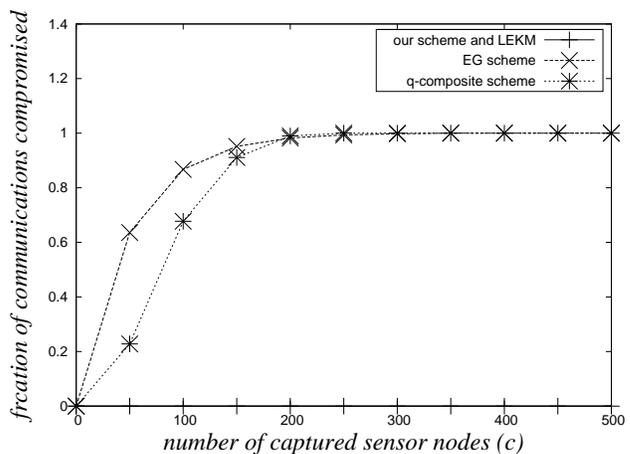}
\caption{Comparison of resilience against node capture among our scheme, the EG scheme, the $q$-composite scheme, and LEKM.} \label{Figure: }
\end{figure}

\begin{figure}[h]
\centering
\includegraphics[scale=0.33,angle=270]{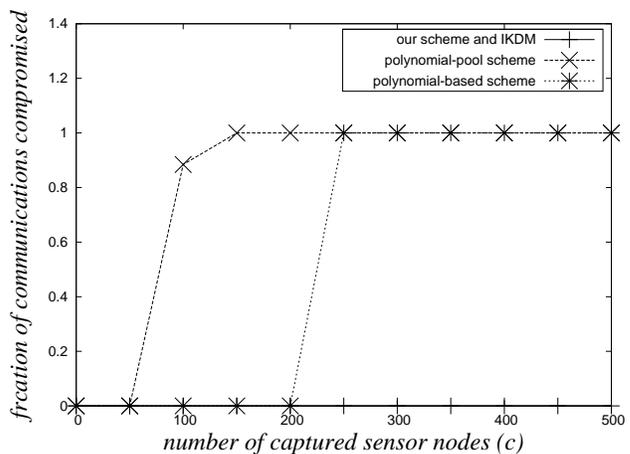}
\caption{Comparison of resilience against node capture among our scheme,
the polynomial-based scheme, the polynomial-pool based scheme, and IKDM.} \label{Figure: }
\end{figure}
\par
\indent The comparison of resilience against sensor node capture between our scheme, the polynomial-based key distribution scheme~\cite{cf:08}, the polynomial-pool based key distribution scheme~\cite{jr:12}, the EG scheme~\cite{cf:01}, the $q$-composite scheme~\cite{cf:02}, the low-energy key management
scheme (LEKM)~\cite{cf:24} and the improved key distribution mechanism (IKDM)~\cite{jr:13} are shown in Figures 6 and 7. We assume that each sensor node is capable of holding $200$ cryptographic keys in its key ring. In LEKM and IKDM, we have taken $100$ clusters and we assume that each cluster has $100$ sensors, since all the sensors will directly communicate to their group head only. The network connectivity for all schemes is taken $\approx 1.00$ with suitable choice of their respective parameters. We note from these figures that even if the number of captured sensor nodes is small, the EG scheme, the $q$-composite scheme, the polynomial-based scheme and the polynomial-pool based scheme reveal a large fraction of total secure communication between non-compromised sensor nodes in the network. We also see that our scheme, LEKM and IKDM provide unconditional security against sensor node capture. Since in our scheme a deployment group can have $221$ members including a group head (an H-sensor node), our scheme supports large-scale network than LEKM and IKDM with the same number of cluster heads (group heads). As a result, though LEKM and IKDM provide unconditional security against sensor node capture, they can not still support a large network as compared to our scheme with the same number of cluster heads (group heads).

\begin{figure}[htbp]
\centering
\includegraphics[scale=0.33,angle=270]{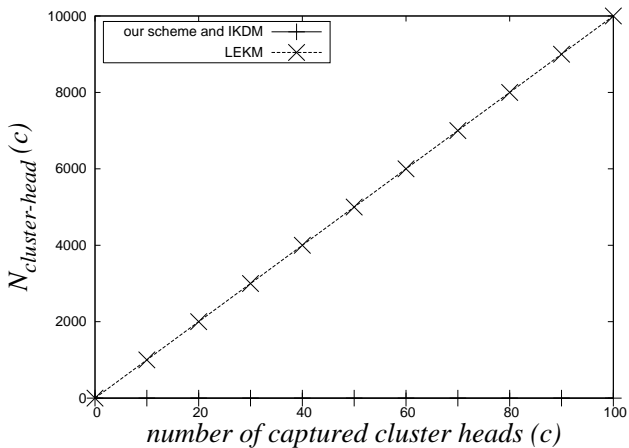}
\caption{Number of compromised regular sensor keys versus number of the compromised cluster heads (group heads) in the network initialization phase. Here $N_{cluster-head} (c)$ denotes the number of compromised keys in sensor nodes after capturing $c$ cluster heads (group heads).} \label{Figure: }
\end{figure}

\par
\indent Figure 8 shows the number of compromised sensor keys vs. number of the compromised cluster heads (group heads) during the network initialization phase. In LEKM and IKDM, we assume that there are $100$ sensors in each cluster and $100$ cluster heads in a network so that they can support $10,000$ sensor nodes. In these schemes, all the sensor nodes will communicate with the cluster head node in a cluster directly.  Since in our scheme, a deployment
group can have $221$ members including a group head (an H-sensor node), our scheme supports $22,000$ regular sensor nodes. In LEKM, any single cluster head's capture could compromise the $100$ sensors' secret keys. From this figure, we note that no matter how many cluster heads (group heads) are compromised in the network initialization phase, our scheme and IKDM provide perfect resilience against cluster head (group head) capture attack. However, in LEKM, as the number of compromising cluster heads increases the number of compromised sensor keys also increases. Thus, we see that our scheme as well as IKDM provide better security against cluster head (group head) capture attack as compared to that for LEKM during network initialization phase. But when the group heads are captured after network initialization phase, all the keys in sensors are compromised in case of LEKM and IKDM. Also, recently Paterson et al.~\cite{cf:29} presented two attacks on IKDM. They showed that their attacks can result in the compromise of most if not all of the sensor node keys after a small number of cluster heads are compromised. In our scheme, only the keys of neighboring sensors of a group head will be compromised. Thus, other sensors will be non-compromised even the group head is compromised. Hence, our scheme provides significantly better security against cluster heads (group heads) capture as compared to that for LEKM and IKDM.

\section{Conclusion}
In this paper, we have proposed an energy-efficient probabilistic group-based key distribution scheme for a large-scale heterogeneous wireless sensor
network. Our scheme always guarantees that any two non-compromised nodes in a deployment group can communicate each other with $100\%$ secrecy. Moreover,
it provides significantly better security against sensor node capture as compared to that for the existing related schemes. Overall, we conclude that our scheme has a better trade-off among network connectivity, security, communication and computational overheads than the existing related schemes. In addition, our scheme supports dynamic regular sensor node addition as well as dynamic group head addition after initial deployment in the network.

\bibliographystyle{IEEEtran}
\bibliography{sensor1}

\begin{thebibliography}{10}
\providecommand{\url}[1]{#1}
\csname url@rmstyle\endcsname
\providecommand{\newblock}{\relax}
\providecommand{\bibinfo}[2]{#2}
\providecommand\BIBentrySTDinterwordspacing{\spaceskip=0pt\relax}
\providecommand\BIBentryALTinterwordstretchfactor{4}
\providecommand\BIBentryALTinterwordspacing{\spaceskip=\fontdimen2\font plus
\BIBentryALTinterwordstretchfactor\fontdimen3\font minus
  \fontdimen4\font\relax}
\providecommand\BIBforeignlanguage[2]{{%
\expandafter\ifx\csname l@#1\endcsname\relax
\typeout{** WARNING: IEEEtran.bst: No hyphenation pattern has been}%
\typeout{** loaded for the language `#1'. Using the pattern for}%
\typeout{** the default language instead.}%
\else
\language=\csname l@#1\endcsname
\fi
#2}}

\bibitem{jr:01}
R.~L. Rivest, A.~Shamir, and L.~M. Adleman, ``A method for obtaining digital
  signatures and public-key cryptosystems,'' \emph{Communications of the ACM},
  vol.~21, pp. 120--126, 1978.

\bibitem{jr:02}
W.~Diffie and M.~E. Hellman, ``New directions in cryptography,'' \emph{IEEE
  Transactions on Information Theory}, vol.~22, pp. 644--654, 1976.

\bibitem{jr:03}
T.~ElGamal, ``A public key cryptosystem and a signature scheme based on
  discrete logarithms,'' \emph{IEEE Transactions on Information Theory},
  vol.~31, pp. 469--472, July 1985.

\bibitem{bk:04}
D.~R. Stinson, \emph{Cryptography Theory and Practice}, 3rd~ed.\hskip 1em plus
  0.5em minus 0.4em\relax Chapman \& Hall/CRC, 2006.

\bibitem{cf:28}
R.~L. Rivest, ``{The RC5 Encryption Algorithm},'' in \emph{Proceedings of the
  second International Workshop on Fast Software Encryption}, vol. 1008, 1994,
  pp. 86--96.

\bibitem{jr:06}
I.~F. Akyildiz, W.~Su, Y.~Sankarasubramaniam, and E.~Cayirci, ``Wireless sensor
  networks : A survey,'' \emph{Computer Networks}, vol.~38, no.~4, pp.
  393--422, 2002.

\bibitem{cf:01}
L.~Eschenauer and V.~D. Gligor, ``A key management scheme for distributed
  sensor networks,'' in \emph{the 9th ACM Conference on Computer and
  Communication Security}, November 2002, pp. 41--47.

\bibitem{cf:02}
H.~Chan, A.~Perrig, and D.~Song, ``Random key predistribution schemes for
  sensor networks,'' in \emph{IEEE Symposium on Security and Privacy},
  Berkeley, California, 2003, pp. 197--213.

\bibitem{cf:04}
D.~Liu and P.~Ning, ``Establishing pairwise keys in distributed sensor
  networks,'' in \emph{Proceedings of 10th ACM Conference on Computer and
  Communications Security (CCS)}, Washington DC, October 27-31 2003, pp.
  52--61.

\bibitem{cf:05}
W.~Du, J.~Deng, Y.~S. Han, S.~Chen, and P.~K. Varshney, ``A key management
  scheme for wireless sensor networks using deployment knowledge,'' in
  \emph{23rd Conference of the IEEE Communications Society (Infocom'04)}, Hong
  Kong, China, March 21-25 2004.

\bibitem{cf:07}
W.~Du, J.~Deng, Y.~S. Han, and P.~K. Varshney, ``A pairwise key
  pre-distribution scheme for wireless sensor networks,'' in \emph{ACM
  Conference on Computer and Communications Security (CCS'03)}, Washington DC,
  USA, October 27-31 2003, pp. 42--51.

\bibitem{jr:10}
D.~Liu and P.~Ning, ``Improving key pre-distribution with deployment knowledge
  in static sensor networks,'' \emph{ACM Transactions on Sensor Networks},
  vol.~1, no.~2, pp. 204--239, 2005.

\bibitem{cf:24}
G.~Jolly, M.~Kuscu, P.~Kokate, and M.~Yuonis, ``A low-energy key management
  protocol for wireless sensor networks,'' in \emph{Proceedings of the Eighth
  IEEE International Symposium on Computers and Communication (ISCC'03)},
  Kemer-Antalya, Turkey, June 30 - July 3 2003.

\bibitem{jr:13}
Y.~Cheng and D.~Agrawal, ``An improved key distribution mechanism for
  large-scale hierarchical wireless sensor networks,'' \emph{Ad Hoc Networks
  (Elsevier)}, vol.~5, no.~1, pp. 35--48, 2007.

\bibitem{cf:18}
A.~K. Das, ``{A Key Reshuffling Scheme for Wireless Sensor Networks},'' in
  \emph{International Conference on Information Systems Security (ICISS 2005),
  Lecture Notes in Computer Science (LNCS)}, vol. 3803, 2005, pp. 205--216,
  {Springer-Verlag}.

\bibitem{cf:08}
C.~Blundo, A.~D. Santis, A.~Herzberg, S.~Kutten, U.~Vaccaro, and M.~Yung,
  ``Perfectly-secure key distribution for dynamic conferences,'' in
  \emph{Advances in Cryptology- CRYPTO'92, LNCS 740}, Berlin, August 1993, pp.
  471--486.

\bibitem{bk:03}
F.~B. Hildebrand, \emph{Introduction to Numerical Analysis}, 2nd~ed.\hskip 1em
  plus 0.5em minus 0.4em\relax New York: Dover, 1974.

\bibitem{jr:12}
D.~Liu, P.~Ning, and R.~Li, ``Establishing pairwise keys in distributed sensor
  networks,'' \emph{ACM Transactions on Information and System Security},
  vol.~8, no.~1, pp. 41--77, 2005.

\bibitem{cf:25}
D.~Huang, M.~Mehta, D.~Medhi, and L.~Harn, ``{Location-aware key management
  scheme for wireless sensor networks},'' in \emph{Proceedings of the 2nd ACM
  workshop on Security of ad hoc and sensor networks SASN '04}, 2004, pp. 29 --
  42.

\bibitem{cf:27}
D.~Liu, P.~Ning, and W.~Du, ``{Group-Based Key Pre-Distribution in Wireless
  Sensor Networks},'' in \emph{Proceedings of 2005 ACM Workshop on Wireless
  Security (WiSe 2005)}, September 2005.

\bibitem{cf:26}
A.~K. Das and I.~Sengupta, ``An effective group-based key establishment scheme
  for large-scale wireless sensor networks using bivariate polynomials,'' in
  \emph{3rd International Conference on Communication Systems Software and
  Middleware (COMSWARE 2008)}, 2008, pp. 9--16.

\bibitem{ms:03}
C.~T. Inc., ``Wireless sensor networks,'' http://www.xbow.com.

\bibitem{jr:08}
O.~Goldreich, S.~Goldwasser, and S.~Micali, ``How to construct random
  functions,'' \emph{Journal of the ACM}, vol.~33, no.~4, pp. 792--807, October
  1986.

\bibitem{jr:17}
S.~Zhu, S.~Setia, and S.~Jajodia, ``{LEAP+: Efficient Security Mechanisms for
  Large-Scale Distributed Sensor Networks},'' \emph{ACM Transactions on Sensor
  Networks}, vol.~2, no.~4, pp. 500--528, November 2006.

\bibitem{cf:09}
D.~Liu and P.~Ning, ``Location-based pairwise key establishments for static
  sensor networks,'' in \emph{ACM Workshop on Security in Ad Hoc and Sensor
  Networks (SASN '03)}, October 2003, pp. 72--82.

\bibitem{cf:29}
\BIBentryALTinterwordspacing
M.~B. Paterson, R.~Holloway, and D.~R. Stinson, ``{Two attacks on a sensor
  network key distribution scheme of Cheng and Agrawal},'' in \emph{Cryptology
  ePrint Archive}, 2008, report 2008/326. [Online]. Available:
  \url{http://eprint.iacr.org/2008}
\BIBentrySTDinterwordspacing

\end{thebibliography}
\end{document}